# In-situ acoustic-based analysis system for physical and chemical properties of the lower Martian atmosphere

F.A. Farrelly, A. Petri, L. Pitolli, G. Pontuale*

CNR – Istituto di Acustica "O.M. Corbino", Roma, Italy

## ABSTRACT

The Environmental Acoustic Reconnaissance and Sounding experiment (EARS), is composed of two parts: the Environmental Acoustic Reconnaissance (EAR) instrument and the Environmental Acoustic Sounding Experiment (EASE). They are distinct, but have the common objective of characterizing the acoustic environment of Mars.

The principal goal of the EAR instrument is "listening" to Mars. This could be a most significant experiment if one thinks of everyday life experience where hearing is possibly the most important sense after sight. Not only will this contribute to opening up this important area of planetary exploration, which has been essentially ignored up until now, but will also bring the general public closer in contact with our most proximate planet.

EASE is directed at characterizing acoustic propagation parameters, specifically sound velocity and absorption, and will provide information regarding important physical and chemical parameters of the lower Martian atmosphere; in particular, water vapor content, specific heat capacity, heat conductivity and shear viscosity, which will provide specific constraints in determining its composition. This would enable one to gain a deeper understanding of Mars and its analogues on Earth.

Furthermore, the knowledge of the physical and chemical parameters of the Martian atmosphere, which influence its circulation, will improve the comprehension of its climate now and in the past, so as to gain insight on the possibility of the past presence of life on Mars. These aspect are considered strategic in the contest of its exploration, as is clearly indicated in NASA's four main objectives on "Long Term Mars Exploration Program" (http://marsweb.jpl.nasa.gov/mer/science).

*Corresponding author.

E-mail address: pontuale@idac.rm.cnr.it

## 1. Introduction

No direct information about the acoustic environment on Mars' surface is available at present and the EAR instrument will therefore investigate a fresh field of study. Sound can be generated by meteorological activity; in particular wind, sandstorm and lighting during these as well as little Martian tornadoes referred to as "dust devils" will most likely be a major source of acoustic. Aeroacoustic sounds associated with the Lander and Rover as well as those related to their activities will be major acoustic sources, due to their proximity to the microphone. However, the most interesting sounds may be those that are not easily predicable and some way of detecting these must be found. In regard to meteorological phenomena, there is much to be gained by the efforts made to study these on Earth once the differences in their acoustic propagation parameters are taken into account.

Regarding the characterization of the acoustic propagation properties of the Martian atmosphere, EASE will adapt standard measurement techniques to the particular requirements of the mission. The physical and chemical properties of the atmosphere of Mars are based on information gained from numerous unmanned space missions. As carbon dioxide is the main component of Mars' atmosphere, the fact that the presence of small quantities of water vapor significantly modifies the acoustic propagation parameters is important, as this enables a sensitive mechanism for the measurement of this fundamental compound.

**2. Scenario**

Acoustic measurements in the context of space planetary exploration were proposed as early as 1966; their concept was to measure the sound speed in a temperature-controlled spiral duct, thus constraining the relative molecular mass and ratio of specific heats. The application in mind was to resolve the relative abundance of $CO_2$, $N_2$, Ar in the then-unknown Martian atmosphere.

More recently, the first acoustic sensors for the outer planets have been sent on their way as part of the Cassini/Huygens mission. The Cassini spacecraft will carry the ESA Huygens probe to Saturn's moon Titan. One of the probe's experiments carries the Surface Science Package (SSP) which includes a speed-of-sound velocimeter and an acoustic sounder (Zarnecki, 1997). This sounder, acting as a SODAR (SOund Detection And Ranging), will be used to infer the landscape roughness around the landing site prior to touch down. Additionally, it will constrain the depth of any liquid hydrocarbon deposit the probe might land in. The Huygens Atmospheric Structure Instrument (HASI) carries a passive microphone (ACU) to search for acoustic events due to possible thunder (Fulchignoni, 1997).

Passive acoustic sensors have been also flown on Soviet Venus landers, with the intent of detecting thunder. Although events were detected, it is unclear whether these were indeed due to thunder or merely aeroacoustic noise caused by the turbulent airflow over the probe. Data from these sensors were used in order to place constraints on surface windspeeds. For more detailed information see the review by Lorenz (1998).

More recently, a microphone was launched towards Mars on the unsuccessful Mars Polar Lander. This purely passive instrument was principally intended for listening to unexpected sounds from another world. It was characterized by confined resources, and thus was strongly limited both in how much sound it could record, and also in the fidelity of those recordings. An improved version of this instrument has been proposed in the framework of the French Net Lander mission, which is planned to be launched in 2007 (http://sprg.ssl.berkeley.edu/marsmic/team.html).

This, to our knowledge, is the current state of the acoustic research in planetary exploration. It is evident enough that, despite their limited resource requirements, the potential of acoustic-based techniques in planetary exploration have not yet been exploited.

### 3. EASE - Environmental Acoustic Sounding Experiment

*3.1 Instrument overview*

The basic structure of the EASE experiment, whose functional block diagram is shown in fig. 4, is composed of two main parts: the acoustic wave guide and the experiment controller and data treatment unit. The cylindrical wave guide (length: 22 cm; diameter: 3 cm) is bounded at the two sides with the transmitting and receiving transducers, although the dimension of the cavity are still somewhat preliminary.

The goal of the EASE experiment is to measure acoustic propagation parameters so that physical and chemical properties of the lower Martian atmosphere can be constrained with the help of established models. That is, by directly determining the wave velocity and absorption as a function of frequency and temperature, the properties of the gas composition can be inferred. Specifically, by analyzing absorption as a function of frequency, molecular relaxation processes can be resolved thus allowing $H_2O$ vapor content to be obtained; similarly, viscosity and thermal conductivity of the gas composition are also determined. Furthermore, by analyzing sound velocity as a function of frequency, the specific heat ratio can be determined. As a result it will possible to verify the compatibility among various proposed atmospheric models.

*3.2 Acoustic propagation parameters*

In the case of monofrequency motion the wave equation is reduced to the Helmholtz equation: $\nabla^2 \mathbf{p} + \mathbf{k}^2 \mathbf{p} = 0$ with $\mathbf{k}$ being the complex wavenumber and $\mathbf{p}$ the acoustic pressure field. The solution to this equation for a plane wave traveling in the +x direction is $\mathbf{p} = P_0 e^{-\alpha x} e^{i(\omega t - kx)}$ where $P_0$ is the maximum acoustic pressure, $\alpha$ is the attenuation coefficient, $\omega$ is the angular frequency and the complex wavenumber is resolved into its components: $\mathbf{k} = k - i\alpha$.

In general the attenuation coefficient takes into account absorption processes, losses due to geometrical attenuation and those due to scattering. The latter two are not applicable to the experimental configuration here described as propagation occurs in a confined acoustically homogenous medium, whereas absorption processes are subject to two types of mechanisms, those due to losses in the medium and those associated with losses at the boundaries of the medium (wall effects). Absorption processes occur when acoustic energy is transferred into other energy forms; ultimately all acoustic energy is degraded into thermal energy (Kinsler, 2000). A sudden change of local pressure conditions will augment the internal energy of the system. This excess energy gained is lost exponentially in time by absorption processes, each of which is characterized by a *relaxation time* $\tau$. Losses that occur in the medium through relaxation mechanisms are divided into three basic types: viscous, heat conduction and those associated with molecular relaxation processes. The general form of the absorption coefficient can be expressed as:

$$\alpha(\omega) = \frac{\omega}{c} \frac{1}{\sqrt{2}} \left[ \frac{\sqrt{1+(\omega\tau)^2} - 1}{1+(\omega\tau)^2} \right]^{1/2} \quad (1)$$

where $c$ is the sound velocity and $\tau$ is the relaxation time (Hill, 1986). Viscous losses result whenever there is a relative motion between adjacent portions of medium, such as during compression and expansion that accompany the propagation of a sound wave. Heat conduction losses result from the conduction of thermal energy between higher temperature condensations and lower temperature rarefactions. The relaxation times associated to these phenomena are typically in the range of $10^{-10}$ s. According to the theory developed by Stokes and Kirchhoff, the so-called *classical absorption* equation is derived from eq.1 by applying the constraint $\omega\tau \ll 1$, so as to obtain:

$$\alpha_{CL}(\omega) \cong \frac{\omega^2}{2\rho_0 c^3} \left[ \frac{4}{3}\eta + (\gamma - 1)\frac{K}{C_P} \right] \quad (2)$$

where $C_p$ is the specific heat at constant pressure and $\rho_0$ is the steady-state density, $\eta$ is the coefficient of shear viscosity and $K$ is the thermal conductivity.

Molecular relaxation processes can also lead to absorption and include the conversion of kinetic energy of the $CO_2$ molecule into its various degrees of freedom (Bass, 1972, Vesovic, 1990, Trengove, 1987); for this molecule, the internal vibrational states have the longest relaxation time ($\tau_M$) enabling this process to be resolved acoustically at low ultrasonic frequencies. Specifically, measurements for dry pure $CO_2$ at 20°C, confirm that for this process the *transition frequency* ($1/2\pi\tau_M$) is about 30 kHz. Water vapor content in $CO_2$ has a considerable influence on this frequency: a 1% concentration of $H_2O$ raises this frequency to 2MHz. This is not directly due to its presence but through its indirect action which lowers the average number of collisions required to transfer energy to the vibrational states of the carbon dioxide molecule, thus acting as a catalyzer. The absorption coefficient $\alpha_M$ is given by eq. 1 in which $\tau$ is replaced by $\tau_M$.

The expected value for transition frequency in Martian atmosphere, relative to water vapor content estimates (~1 $^0/_{00}$) and temperature values (~220 K), may be reasonably located inside EASE frequency range (Petropoulos, 1989). Because relaxation time decreases with increasing humidity and does so most rapidly when the gas is almost dry this instrument is expected to be particularly sensitive to $H_2O$ vapour content. The actual value for the relaxation time depend not only on temperature, pressure and $H_2O$ concentration but it is also affected by the presence of other gases (Shields, 1967). It is therefore necessary to perform specific experiments in order to determine $\tau_M$ as a function of these parameters.

Wall effects in the low frequency range are the principal source of sound energy absorption associated with viscous resistance offered to fluid motion at the walls of the cavity and the exchange of heat energy between the walls and the fluid. When these effects dominate, their contribution to the absorption coefficient is:

$$\alpha_W = \frac{1}{rc}\sqrt{\frac{\eta_e \omega}{2\rho_0}} \quad ; \quad \eta_e = \eta\left[1 + (\gamma - 1)\sqrt{\frac{K}{C_p \gamma \eta}}\right] \qquad (3)$$

Where $r$ is the radius of the cavity and $\eta_e$ is the *effective viscosity* (Morse, 1968).

The complete absorption coefficient is therefore given by the sum:

$$\alpha(\omega) = \alpha_{CL}(\omega) + \alpha_M(\omega) + \alpha_W(\omega) \qquad (4)$$

This theoretical expression is shown as solid line curve in fig.1, where the effects of the various contributions are clearly distinguishable; in particular the logarithmic scales show a quadratic dependency on the frequency in the classical (dashed line) and molecular (dot-dashed line) relaxation regions when their approximated form is valid, and the square root dependency in the region where the wall effect dominates. Preliminary experiments of an acoustic cavity (see fig. 2) have been conducted in a custom built simulation chamber capable of reproducing the pressure and temperature characteristics of the lower Martian atmosphere. The vacuum facility is composed of a rotary pump and condenser-type pressure meter, while the temperature is controlled by immersing the simulation chamber in a melting mixture of distilled water and anti-freeze. Two thermocouples placed inside the chamber are used to measure temperature. Some test runs have been conducted in physical conditions close to those expected "in-situ" and show promising results (see fig. 3).
In particular, both transmitting and receiving transducers have functioned properly, exhibiting good signal to noise ratio even at low transmission powers, at pressures down to 7 mbar and temperatures of about 240 K.

*3.3 Instrument description and operating modes*

Various techniques can be used to measure the propagation parameters, and they can be classified in two categories; spectral methods in which a continuous monofrequency signal is transmitted, and time-of-flight methods in which a modulated signal is used.

In the spectral method resonance peaks in the acoustic cavity's transmission function are determined by its boundary conditions and the medium's propagation parameters: phase velocity and attenuation. The amplification given to the propagating wave with a frequency near a resonance peak makes this technique suitable for very low power measurements, particularly when a lock-in amplification system is used for detecting the received signal. The discrete set of frequencies which can be studied are determined by the geometry of the wave guide and the phase velocity of the medium.

For the time-of-flight methods, one interesting technique uses a Pseudo Random Noise (PRN) modulated continuous wave signal in order to determine the acoustic propagation parameters: in this mode the signal received after travelling the length of the acoustic cavity is correlated to a delayed version of the transmitted signal code sequence and a peak value of this correlation product is found when the delay ($t_d$) is set to the time of flight of the propagating signal. By changing the carrier frequency of the transmitted signal, its spectral content can be altered to some degree, making it possible to determine the acoustic propagation properties of the medium as a function of frequency. Due to the high noise rejection capabilities of this detection technique, it is possible to implement a low-power system which is both tolerant to electrical noise and produces an insignificant amount of interference.

The current design of the EASE instrument enables several operating modes to be performed allowing in this way to choose between tradeoffs in these. A block diagram of the instrument is shown in fig. 4. Three Direct Digital Synthesis (DDS) wave generators are used; DDS0 for the transmission signal, while the other two are used in the detection phase as part of a lock-in circuit, when operating in monofrequency mode. For PRN measurements the relative phase between DDS0 and the other two is varied until the maximum of the correlation is found.

## 4. EAR - Environmental Acoustic Reconnaissance

*4.1 Instrument overview*

The operative rational of EAR is to deliver as much scientifically valuable information regarding the acoustic environment on Mars, as can be had under all foreseeable conditions. As a secondary objective, it would be nice to know what Mars would "sound" like to a human ear.

One important aspect is determining the frequency range over which the instrument is to operate; infrasonic sounds could be valuable in furnishing information from distant sources as for example meteorites (Williams, 2001), but in order to discriminate between pressure waves caused by gusts of wind and sound propagation, they require large distributed detectors. Strong winds however, which are likely to be found in this case, introduce noise severely degrading signal quality, even when sensors are properly designed (see, for example, www.nemre.nn.doc.gov/review2001/CD/Print/07-06.pdf). This makes infrasound measurements difficult to perform given mission constraints and the local environment. As frequency increases, attenuation increases sharply as is described in the section dedicated to the EASE instrument, so one can expect that only nearby sources (e.g. elements of the mission itself) will contribute to the higher end of the audio frequency spectrum (10 kHz ca.).

In planetary exploration missions, instruments contend over a pool of constrained resources, the most critical of which in EAR's case are telemetry and power while the requirements on mass and volume are relatively moderate. The power and telemetry pair are in someway tied together, as sophisticated "in-situ" analysis requires complex calculations which are power intensive but can in return dramatically reduce telemetry requirements.

To achieve this, a self-adaptive scheme shall be implemented in such a way as to be able to hand over the best available data over a wide range of telemetry rates; even if telemetry resources should be greatly reduced due to say, transmission fault conditions, EAR must deliver whatever valuable information it can, given its operating constraints. On the other hand it should be able to take advantage of favorable conditions, should these come up, to hand-over as much data as can be handled. In fact, considering what little knowledge there is of the acoustic environment of Mars, it would be interesting to have continuous monitoring of a large spectrum of frequencies, from multiple transducers. However, technical restraints must be considered; even using the most sophisticated compression techniques telemetry requirements for such an experiment would be quite substantial. A single CD quality channel (16 bit resolution and over 40 ks/s) would generate, with a 4:1 compression ratio (see below), almost 2 Mbits for a 10 second sample. It is therefore essential to provide a mechanism for selecting acoustic events, so that the most promising ones will be transmitted with higher priority and compress the data so that redundant information content is reduced to a minimum.

*4.2 Instrument description*

The EAR instrument is composed of two subsystems (see fig. 5); the Microphone and Mounting (MM), and the Basic Reconnaissance Analysis and Interface Node (BRAIN) subsystem, where the main electronics system is housed. The MM subsystem will be suitably located so as to have as little acoustic interference as possible on the part of the lander and its mission packages. The microphone itself will probably be a preamplified electret and will have an appropriate wind-shielding device. Electret microphones have a frequency response (~20÷18,000 Hz) and sensitivity (~50 mV/Pa) that are adequate for the purposes of this instrument and require a far lower polarization voltage than condenser and piezoelectric types of microphones enabling the use of simpler electronics.

The BRAIN subsystem will filter the analogue signal and digitise it (conceivably with a 16 bit resolution at 40 kilo-samples per second), identify potential acoustic events and store them in a workspace for further processing. Statistical data will be gathered giving general characteristics for long term coverage, and this concise information will be imparted with high priority.

*4.3 Operational Modes*

To improve the selection criteria it would be useful to take into account all additional information available regarding lander/rover activities, local meteorological conditions, nighttime/daytime operations, etc., during the acoustic acquisition periods, so as to operate in the most appropriate experiment profile. Because of the uncertainties associated with hearing unexpected sound sources, the availability of near real-time command (24-36 hour delay times) operations may be particularly useful for the EAR instrument so that, for example, priority hierarchies may be modified. Various types of acoustic events shall be available for handling; e.g., along with transformed time series data, statistical data concerning average power spectrum will be collected and treated as specific types of acoustic events. This valuable information will be very concise and will have high dispatch priority. The mechanism foreseen, which shall be in charge of providing the proposed self-adapting telemetry scheme, will involve subdividing the acquired acoustic signals into data events, classifying and attributing a dispatch priority level to them. This way scientifically important data, with low data transmission requirements shall have the best chance of being delivered, whereas events with less chance of getting through tight transmission bottlenecks will have a lower dispatch priority. This adaptive dispatch scheme requires a memory management system to deal with contrasting memory requirements. A similar priority system will be implemented to enforced the planned mission modes and ensure optimal scientific data availability for the scientific effort.

*4.4 In Situ Analysis*

EAR can potentially implement sophisticated computations so that the available telemetry can be used efficiently. Lossy time series compression is the most promising technique to deliver rich data to earth for further analysis; essentially this entails choosing which aspect in the original signal is to be preserved most faithfully. A very popular form of compression is that based on psychoacoustic analysis in which the properties of the Human ear are taken into account so that the listener does not notice a significant difference when hearing the compressed version as compared to the original one. The most significant result of these algorithms is to eliminate frequency components masked by a nearby robust frequency component. This may severely affect the quality of the information content of the signal so other techniques should also be investigated.

Time-series data, subdivided into acoustic events, will be transformed into wavelet or Fourier space so as to perform functions similar to undersampling with data-loss compression techniques and to aid in the event selection and prioritization process.

In fig. 6 a time series signal (a) and its Short Time Fast Fourier Transfer (STFFT) (b) are shown as an example of an acoustic event. In STFFT the signal is subdivided into short time slots which are then decomposed into a set of equally spaced frequency bins, whereas in wavelet transformations the signal is processed by constant-Q filters (Van den Berg, 1999), so that bins are spaced in constant frequency ratios. The local nature of wavelet functions is an advantage when relatively short duration signals are analyzed. Data loss compression techniques based on wavelet compression typically keep, once the data has been transformed, only the strongest components with the highest resolution while weaker coefficients are either represented in a coarser fashion or discarded altogether (Press, 1996). Upon reconstruction through the inverse wavelet transformation, a modified signal similar to the original may be had. In order to establish to what extent a signal has been degraded after such a process, an appropriate index of the information content should be determined. One such index is the energy content of the signal. Fig.7 (a) shows an example of how this index can be used; the ratio of the energy content of the modified signal to that of the original one (E) is plotted versus the ratio of the number of coefficients used to their original number (n). In part (b) of the figure, a portion of the original (gray) and modified (black) time series signal is shown in which 12.5% of the coefficients have be kept, the rest having been set to 0, prior to reconstruction. The principal detriment appears to have a local nature and set in a weaker portion of the original signal. Techniques such as this may be useful also in the identification of the acoustic events and their prioritization, which will be mandatory due to the limitations in telemetry rate which impose a severe restraint on number, length and bandwidth of the sound samples available for further analysis on Earth. If one cannot be totally sure that these samples will all be sent back, then it is wise to establish a priority queue for their transmission. To obtain as wide a selection of sounds as possible, the criteria used in establishing the priority of the acoustic events should be varied. The actual criteria will be established through testing between simulated and acquired sounds. Possible selection mechanisms can be very simple or quite sophisticated. The Mars-Mic instrument, for example, chooses the loudest sound splice during the listening phase prior receiving the send-sound telecommand. A more complex selection can be had performing a correlation of the acoustic event with a set of sounds so as to reject or select sound with specific characteristics. This can be done quite efficiently by transforming the sound splice into Fourier space and multiplying each coefficient by that of the response function stored directly in Fourier space. However, in this way unexpected sounds may not crop-up. Clustering in time-frequency space, a technique used typically in image processing, may potentially be used to select sounds which have a sufficient time-bandwidth product.

5. Conclusions

The two instruments of the EARS experiment have been described on a baseline configuration both in theoretical rationale and practical implementation. For the EAR experiment possible "in situ" analyses have been described indicating some data reduction schemes and event selection criteria. The sound propagation parameters measured by EASE experiment, allow the evaluation of various physical and chemical parameters of the Martian atmosphere and could therefore be used to establish fixed points in atmosphere models to be adopted for the Martian environment.

**Acknowledgments**

This work was supported by a grant from Agenzia Spaziale Italiana (ASI).

We would like to thank the reviewers for useful comments and suggestions that greatly improved the paper.